\documentclass[twocolumn,english,superscriptaddress,hidepacs,amsmath,amssymb,prl,floatfix]{revtex4}
\usepackage[T1]{fontenc}
\usepackage[latin9]{inputenc}
\usepackage{graphicx}
\usepackage{amssymb}

\makeatletter

\newcommand*{\ket}[1]{\mathopen{|}#1\mathclose{\rangle}}
\newcommand{\rs}{\rm \scriptscriptstyle }

\makeatother

\usepackage{babel}

\begin{document}

\preprint{APS/123-QED}

\title{A Rydberg Quantum Simulator}

\author{Hendrik Weimer}

\affiliation{Institute for Theoretical Physics III, Universit\"at Stuttgart, 70550
Stuttgart, Germany}

\author{Markus M\"uller}

\affiliation{Institute for Theoretical Physics, University of Innsbruck, and Institute
for Quantum Optics and Quantum Information of the Austrian Academy
of Sciences, A-6020 Innsbruck, Austria}

\author{Igor Lesanovsky}

\affiliation{Institute for Theoretical Physics, University of Innsbruck, and Institute
for Quantum Optics and Quantum Information of the Austrian Academy
of Sciences, A-6020 Innsbruck, Austria}

\affiliation{School of Physics and Astronomy, The University of Nottingham, Nottingham
NG7 2RD, United Kingdom}

\author{Peter Zoller}

\affiliation{Institute for Theoretical Physics, University of Innsbruck, and Institute
for Quantum Optics and Quantum Information of the Austrian Academy
of Sciences, A-6020 Innsbruck, Austria}

\author{Hans Peter B\"uchler}

\affiliation{Institute for Theoretical Physics III, Universit\"at Stuttgart, 70550
Stuttgart, Germany}

\date{\today}

\maketitle

\textbf{Following Feynman and as elaborated on by Lloyd, a universal
quantum simulator (QS) is a controlled quantum device which 
reproduces the dynamics of any other many particle quantum system
with short range interactions. This dynamics can refer to both coherent
Hamiltonian and dissipative open system evolution. Here we show that
laser excited Rydberg atoms in large spacing optical or magnetic lattices
provide an efficient implementation of a universal QS for spin models involving}
\textbf{\emph{(high order)}} \textbf{\emph{n-body interactions}}\textbf{.
This includes the simulation of Hamiltonians of exotic spin models
involving n-particle constraints such as the Kitaev toric code, color
code, and lattice gauge theories with spin liquid phases. In addition, it provides the ingredients for
dissipative preparation of entangled states based on engineering n-particle
reservoir couplings. The key basic building blocks of our architecture
are efficient and high-fidelity} \textbf{\emph{n-qubit entangling
gates via auxiliary Rydberg atoms}}\textbf{, including a possible
dissipative time step via optical pumping. This allows to mimic the
time evolution of the system by a sequence of fast, parallel
and high-fidelity n-particle coherent and dissipative Rydberg gates.} 

Laser excited Rydberg atoms \cite{Gallagher1994,Tong2004,Singer2004,Cubel2005,Vogt2006,Mohapatra2007,Heidemann2007} stored in large spacing optical lattices \cite{Nelson2007} 
or magnetic trap arrays \cite{Whitlock2009} offer unique possibilities for implementing
scalable quantum information processors. In such a setup single atoms
can be loaded and kept effectively frozen at each lattice site, with long-lived
atomic ground states representing qubits or effective spin degrees of freedom.
Lattice spacings of the order of a few $\mu$m allow \emph{single site addressing}
with laser light, and thus individual manipulation and readout of
atomic spins. Exciting atoms with lasers to high-lying Rydberg states and exploiting the strong and long-range dipole-dipole or Van der Waals interactions between Rydberg states provides fast and \emph{addressable 2-qubit entangling operations} or effective
spin-spin interactions; recent theoretical proposals have extended
Rydberg-based protocols towards a single step, high-fidelity entanglement
of a mesoscopic number of atoms \cite{Moller2008,Muller2009}. Remarkably, the basic building blocks
behind such a setup have been demonstrated recently in the laboratory
by several groups \cite{Gaetan2009,Urban2009}.

Motivated by and building on these new experimental possibilities,
we discuss below a Rydberg QS for many body spin models.
As a key ingredient of our setup (see Fig. 1) we introduce additional 
auxiliary qubit atoms in the lattice, which play a two-fold role:
First, they control and \emph{mediate} \emph{effective $n$-body
spin interactions} among a subset of $n$ system spins residing in 
their neighborhood in the lattice. In our scheme this is achieved efficiently 
making use of single-site addressability and a parallelized multi-qubit gate, 
which is based on a combination of strong and long-range Rydberg interactions 
and electromagnetically induced transparency (EIT), as suggested recently in 
Ref. \cite{Muller2009}. Second, the auxiliary atoms can be optically pumped, 
thereby providing a dissipative element, which in combination with Rydberg 
interactions results in \emph{effective collective dissipative dynamics} of a 
set of spins located in the vicinity of the auxiliary particle, which itself eventually 
factors out from the system spin dynamics. The resulting coherent and dissipative 
dynamics on the lattice can be represented by, and thus simulates a master 
equation $\dot{\rho}=-(i/\hbar)\left[H,\rho\right]+{\cal L}\rho$ \cite{Breuer2002},
where the Hamiltonian $H=\sum_{\alpha}H_{\alpha}$ is the sum of $n$-body 
interaction terms, involving a quasi-local collection of spins in the lattice. The 
Liouvillian term ${\cal L}=\sum_{\beta}{\cal D}(c_{\beta})$ with ${\cal D}(c)\rho=
c\rho c^{\dagger}-\frac{1}{2}c^{\dagger}c\rho-\rho\frac{1}{2}c^{\dagger}c$ in 
Lindblad form governs the dissipative time evolution, where the many-particle 
quantum jump operators $c_{\beta}$ involve products of spin operators in a given neighborhood. 

The actual dynamics of our system is performed as a stroboscopic
sequence of coherent and dissipative operations involving
the auxiliary Rydberg atoms over time steps $\tau$, with the master
equation emerging as a coarse grained description of the time evolution. 
For purely coherent dynamics governed by the Hamiltonian, this is the familiar 
{}``digital QS'' \cite{Feynman1982,Lloyd1996} where for each time step the 
evolution operator is implemented via a Trotter
expansion $e^{-iH\tau/\hbar}\approx\prod_{\alpha}e^{-iH_{\alpha}\tau/\hbar}$
and a certain error associated with the non-commutativity of the quasi-local
interactions $H_{\alpha}$. The concept of stroboscopic time evolution is readily 
adapted to the dissipative case by interspersing coherent propagation and dissipative 
time steps $e^{{\cal L}\tau}\approx\prod_{\beta}e^{{\mathcal{D}}(c_{\beta})\tau}$,
providing an overall simulation of the master equation by sweeping
over the whole lattice with our coherent and dissipative operations. Many of these steps 
can in principle be done in a highly parallel way, rendering the time for a simulation step 
independent on the system size. 
In our scheme the characteristic energy scale of the many-body interaction terms is essentially 
the same for two-body, four- or higher-order interaction terms, and mainly limited by the fast 
time-scale to perform the parallel mesoscopic Rydberg gate operations.

\section{Coherent and dissipative many body spin dynamics}

Before proceeding with the concrete physical implementation of our Rydberg QS, 
we find it convenient to discuss  special spin models and master equations of interest, starting with an explicit example: Kitaev's toric code. We will discuss the realization of a more complex setup of a three-dimensional $U(1)$ lattice gauge theory 
giving rise to a spin liquid phase in the last section.

Kitaev's \textit{toric code} is a paradigmatic, exactly solvable model, out of a large class 
of spin models, which have recently attracted a lot of interest in the context of studies on 
topological order and quantum computation. It considers a two-dimensional
setup, where the spins are located on the edges of a square lattice
\cite{Kitaev2003}. The Hamiltonian $H=-E_{0}\left(\sum_{p}A_{p}+\sum_{s}B_{s}\right)$
is a sum of mutually commuting stabilizer operators $A_{p}=\prod_{i\in p}\sigma_{i}^{x}$
and $B_{s}=\prod_{i\in s}\sigma_{i}^{z}$, which describe \emph{four-body
interactions} between spins located around plaquettes ($A_{p}$) and
vertices ($B_{s}$) of the square lattice (see Fig. \ref{fig1}b).
The ground state  of the Hamiltonian is a simultaneous eigenstates of all stabilizer 
operators $A_{p}$ and $B_{s}$ with eigenvalues $+1$,  and gives rise to a topological phase: 
the ground state degeneracy depends on the boundary conditions and topology of the setup, and
the elementary excitation exhibit anyonic statistics under braiding.
The toric code shows two types of excitations  corresponding to $-1$ eigenstates of each 
stabilizer $A_p$ (``magnetic charge'') and $B_p$ (``electric charge'').

A dissipative dynamics which {}``cools'' into the ground state manifold
is provided by a Liouvillian with quantum jump operators,
\begin{equation}
c_{p}=\frac{1}{2}\sigma_{i}^{z}\left[1-A_{p}\right],\hspace{10pt}c_{s}=\frac{1}{2}\sigma_{j}^{x}\left[1-B_{s}\right],
\label{eq:Kitaev_jump_operators}
\end{equation}
with $i\in p$ and $j\in s$, which act on four spins located around plaquettes $p$ and vertices $s$, respectively.
The jump operators are readily understood as operators which {}``pump''
from $-1$ into $+1$ eigenstates of the stabilizer operators: the part $(1-A_{p})/2$ is a projector onto the eigenspace of 
$A_p$ with $-1$ eigenvalue, while all states in the $+1$ eigenspace are dark states. 
The subsequent  spin flip $\sigma_{j}^{x}$  transfers the excitation to the neighboring plaquette.
The jump operators then give rise to a random walk of anyonic excitations on the lattice, and whenever two excitations of the 
same type meet they are annihilated, resulting in a cooling process, see Fig.~\ref{fig3}.
Similar arguments apply to $c_{s}.$ Efficient cooling is achieved by alternating the index $i$ of the spin, which is flipped.

Our choice of the jump operator follows the idea of reservoir engineering
of interacting many-body systems as discussed in Ref. \cite{Diehl2008,Kraus2008}.
In contrast to alternative schemes for measurement based state preparation
\cite{Aguado2008}, here, the cooling is part of the time evolution
of the system. These ideas can be readily generalized to more complex stabilizer
states and to setups in higher dimensions, as in, e.g., the color
code (see Fig. \ref{fig1}c) \cite{Bombin2006}. 
As a final remark we would like to mention that the toric code can also be derived as a perturbative limit of a Hamiltonian with two-body interactions on a honeycomb lattice \cite{Kitaev2006}, of which implementations have been suggested both for cold atoms \cite{Duan2003} and condensed matter systems \cite{Jackeli2009}. In our approach the higher-order interactions arise in a non-perturbative way and the scheme also allows for dissipative state preparation.

\section{Implementation of a single time step}

\begin{figure*}[h!]
\includegraphics[width=0.95\columnwidth]{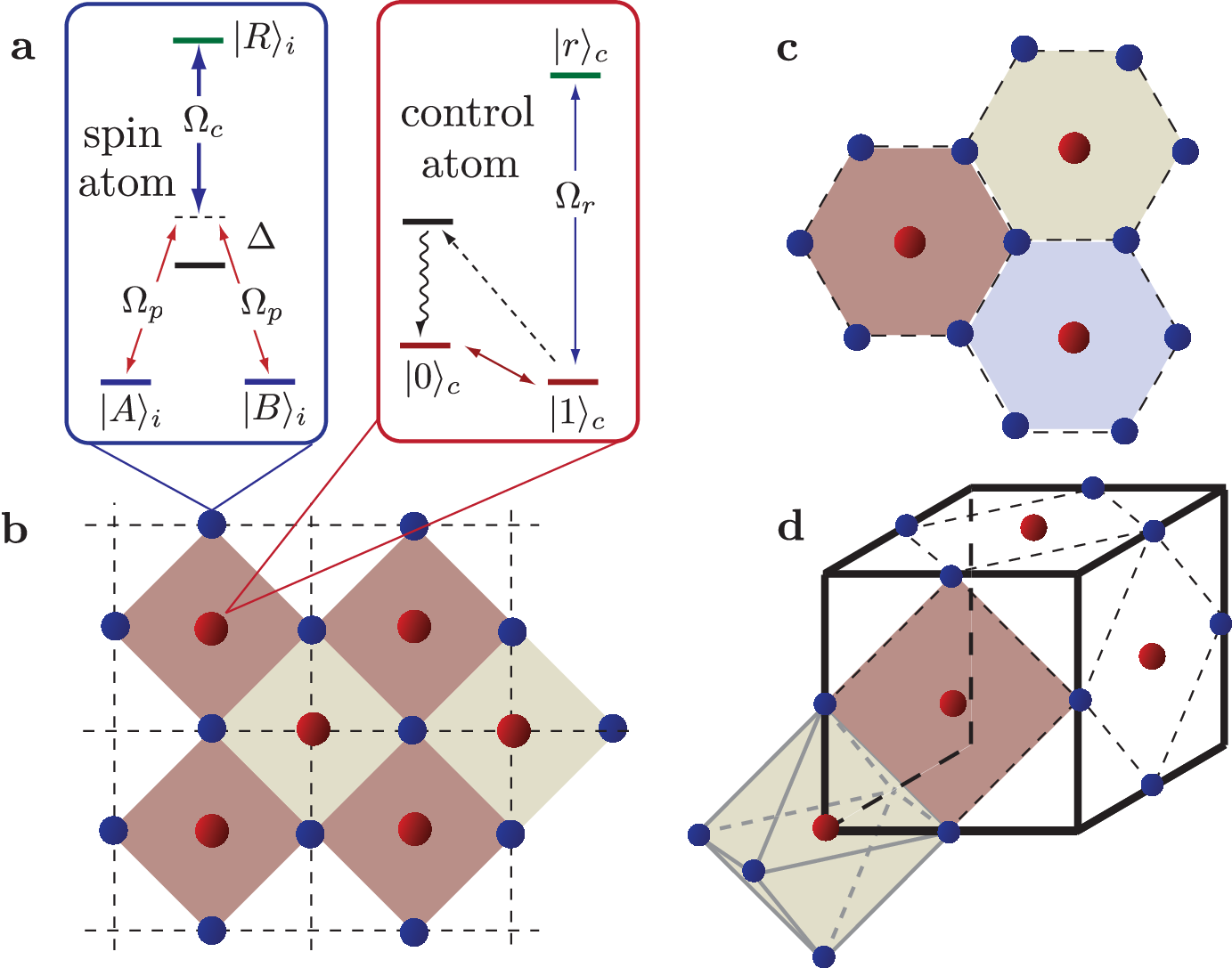} 
\caption{Setup of the system: a) Two internal states $|A\rangle_{i}$ and $|B\rangle_{i}$
give rise to an effective spin degree of freedom. These states are
coupled to a Rydberg state $|R\rangle_{i}$ in two-photon resonance,
establishing an EIT condition. On the other hand, the control atom
has two internal states $|0\rangle_{c}$ and $|1\rangle_{c}$. The
state $|1\rangle_{c}$ can be coherently excited to a Rydberg state
$|r\rangle_{c}$ with Rabi frequency $\Omega_{r}$, and can be optically
pumped into the state $|0\rangle_{c}$ for initializing the control
qubit. b) For the toric code, the system atoms are located on the
links of a 2D square lattice, with the control qubits in the centre
of each plaquette for the interaction $A_{p}$ and on the sites of
the lattice for the interaction $B_{s}$. Setup required for the implementation
of the color code (c), and the $U(1)$ lattice gauge theory (d). }
\label{fig1} 
\end{figure*}

\begin{figure*}[h!]
\includegraphics[width=0.95\columnwidth]{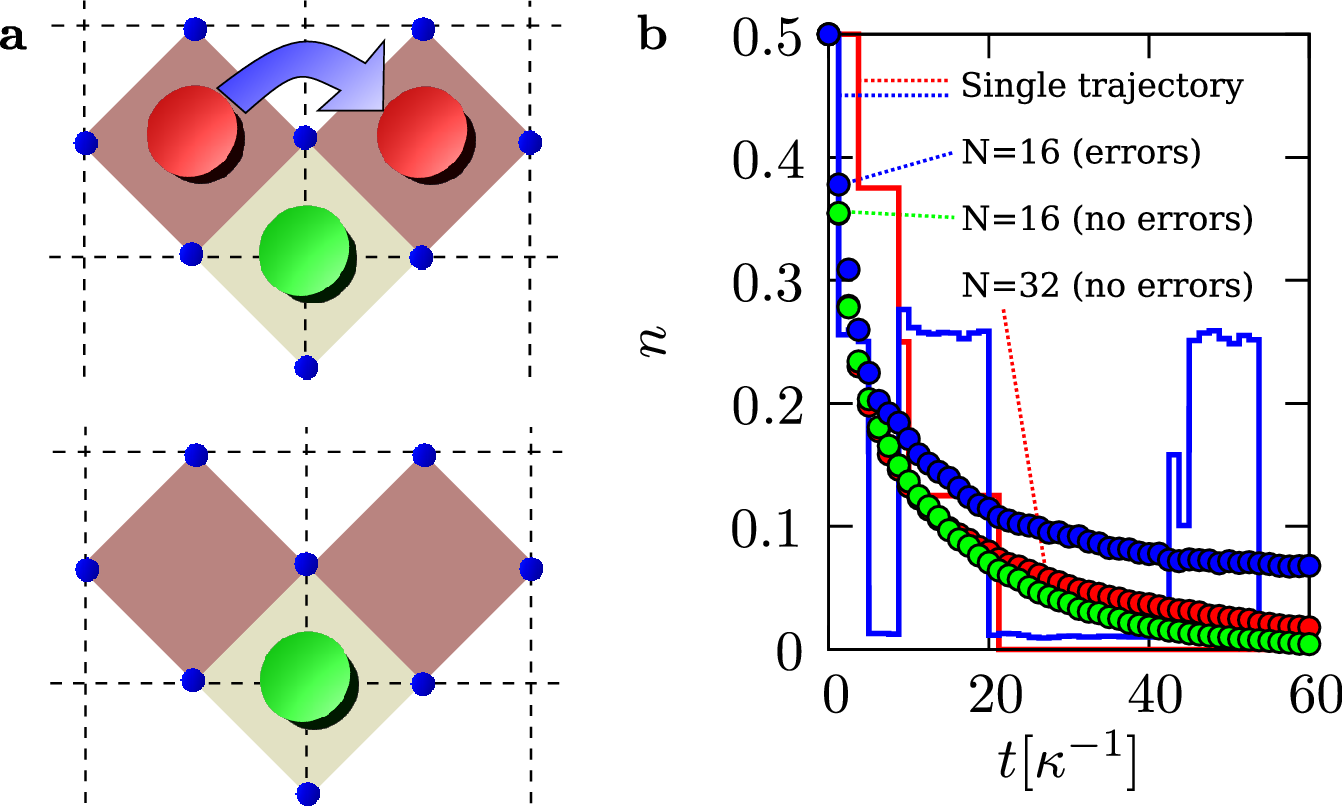} 

\caption{Cooling of the toric code: a) A dissipative time step moves
one anyonic excitation (red dot) on top of a second anyon
sitting on a neighboring plaquette, annihilating each other and 
thus lowering the internal energy of the system. The
anyon of different type (green dot) is unaffected as moves of anyons occur only with a small probability. b) Numerical simulation
of the cooling for $N$ lattice sites (periodic boundary conditions). 
Single trajectories for the anyon density $n$ over time are shown as solid lines. Filled
circles represent averages over 1000 trajectories. The initial state for the simulations is the 
fully polarized, experimentally easily accessible state of all spins down. For perfect gates 
the energy of the system reaches the ground state energy in the long
time limit, while for imperfect gates heating events can occur (blue solid line) and a finite 
density of anyons $n$ remains present (blue circles). 
In this example the phase shift determining the cooling rate was set to $\theta = 1.25$ providing a characteristic time scale
$\kappa^{-1} \sim 8 \mu {\rm s}$, while the parameter quantifying the 
gate error was $|Q|=0.1$ (see the Methods section for details).}

\label{fig3} 
\end{figure*}

We now turn to the physical implementation of the digital quantum simulation. 
The system and auxiliary atoms are stored in a deep optical lattice or
magnetic trap arrays with one atom per lattice site, where the motion of the 
atoms is frozen and the remaining degree of freedom of the system and auxiliary 
atoms are effective spin$-1/2$ systems described by the two long-lived ground 
states $|A\rangle_{i}$ and $|B\rangle_{i}$ and $|0\rangle_c$ and $|1\rangle_c$, 
respectively (see Fig.~\ref{fig1}a). We first discuss the elements of the digital QS for a small local setup, and 
present the extension to the macroscopic lattice system below. To be specific, we will
focus on a single plaquette in the example of Kitaev's toric code outlined above.

\begin{figure*}[h!]
 \includegraphics[width=0.95\columnwidth]{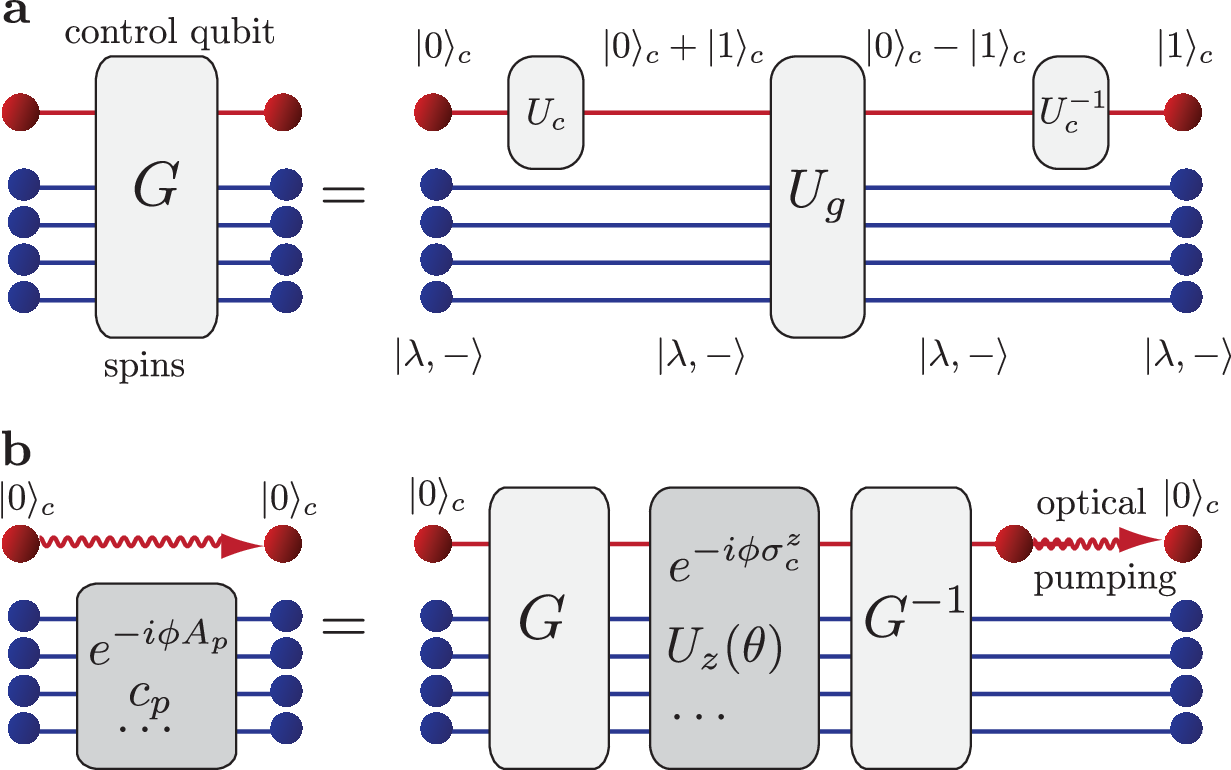} 

\caption{Single time step: a) The gate sequence $G$ coherently maps
the information, whether the system spins reside in any eigenstate
$|\lambda,-\rangle$ ($|\lambda,+\rangle$) corresponding to the eigenvalue
$-1$ ($+1$) of the many-body interaction $A_{p}$ onto the internal
state $|1\rangle_{c}$ ($|0\rangle_{c}$) of the control qubit. b)
After the mapping $G$, we apply gate operations, which depend on
the internal state of the control qubit. Finally, the mapping $G$
is reversed and the control qubit is incoherently reinitialized in
state $|0\rangle_{c}$ by optical pumping. At the end of the complete
sequence the dynamics of the control qubit factors out.
\label{fig2}}
\end{figure*}

The implementation of the four-body spin interaction 
$A_{p}=\prod_{i}\sigma_{i}^{x}$ and the jump operator $c_p$ uses an auxiliary 
qubit located at the centre of the plaquette (see Fig.\ref{fig1}b). The general approach 
then consists of three steps (see Fig.~\ref{fig2}b):
(i) We first perform a gate sequence $G$ which encodes 
the information whether the four spins are in a $+1$ or $-1$ eigenstate of $A_{p}$ in the
two internal states of the auxiliary atom. (ii) In a second step,
we apply gate operations, which depend on the internal state of the
control qubit. Due to the previous mapping these manipulations of
the control qubit are equivalent to manipulations on the subspaces
with fixed eigenvalues of $A_{p}$. (iii) Finally, the mapping $G$ is reversed,
and the control qubit is re-initialized incoherently in its
internal state $|0\rangle$ by optical pumping.

The mapping $G$ is a sequence of three gate operations 
\begin{equation}
G=U_{c}(\pi/2)^{-1}U_{g}U_{c}(\pi/2),
\end{equation}
where $U_{c}(\pi/2)=\exp(-i\pi\sigma_{y}/4)$ is a standard $\pi/2$-single
qubit rotation of the control qubit and the parallelized many-body Rydberg gate \cite{Muller2009} takes the form (see Fig. \ref{fig1}a for the required electronic level scheme and the Methods section for a brief summary)
\begin{equation}
U_{g}=\left|0\right\rangle \!\left\langle 0\right|_{c}\otimes{\bf 1}+\left|1\right\rangle \!\left\langle 1\right|_{c}\otimes\prod_{i \in p} \sigma_{i}^{x}\label{eq:gate}.
\end{equation}
For the control qubit initially prepared in $|0\rangle_{c}$, the gate $G$ coherently transfers the control qubit into the state $|1\rangle_{c}$ ($|0\rangle_{c}$) for any system state
$|\lambda,-\rangle$ ($|\lambda,+\rangle$), with $|\lambda,\pm \rangle$ denoting the eigenstates of $A_{p}$, i.e., $A_{p}|\lambda,\pm\rangle=\pm|\lambda,\pm\rangle$,
see Fig. \ref{fig2}.

For the \textit{coherent time evolution}, the application of a phase
shift $\exp\left(i\phi\sigma_{c}^{z}\right)$ on the control qubit
and the subsequent reversion of the gate, $G^{-1}$, implements the
time evolution according to the many-body interaction $A_{p}$, i.e.,
\begin{equation}
U_{\rs{int}}=\exp\left(i\phi A_{p}\right)=G^{-1}\exp\left(i\phi\sigma_{c}^{z}\right)G.\end{equation}
The control qubit returns to its initial state $|0\rangle_{c}$ after
the complete sequence and therefore effectively factors out from the
dynamics of the system spins. For small phase imprints $\phi\ll1$,
the mapping reduces to the standard equation for coherent time
evolution \begin{equation}
\partial_{t}\rho=-\frac{i}{\hbar}E_{0}\:\left[-A_{p},\rho\right]+o(\phi^{2}).\end{equation}
The energy scale for the four-body interaction $A_{p}$ becomes $E_{0}=\hbar\phi/\tau$
with $\tau$ the time required for the implementation of a single time step.

On the other hand, for the \textit{dissipative dynamics}, we are interested
in implementing the jump operator $c_{p}$ (see Eq. \ref{eq:Kitaev_jump_operators}). To this purpose, after the mapping $G$, we apply a controlled spin flip onto one of the four
system spins,
\begin{equation}
U_{\rs Z,i}(\theta) = \left|0\right\rangle \!\left\langle 0\right|_{c}\otimes{\bf 1}+\left|1\right\rangle \!\left\langle 1\right|_{c}\otimes\Sigma
 \end{equation}
 with $\Sigma = \exp(i \theta \sigma_i^z)$. As desired, the sequence $G^{-1}U_{\rs Z,i}(\theta)G$ leaves
the low energy sector $|\lambda,+\rangle$ invariant since these states are mapped onto $\ket{0}_c$ and are therefore unaffected by $U_{\rs Z,i}(\theta)$. In contrast
- with
a certain probability - the sequence performs a controlled spin flip on the
states $|\lambda,-\rangle$. Once a spin is flipped,
the auxiliary qubit remains in the state $|1\rangle_{c}$, and optical pumping from $|1\rangle_{c}$ to $|0\rangle_{c}$
is required to re-initialize the system, guaranteeing that the control qubit again
factors out from the system dynamics. The optical pumping constitutes the
dissipative element in the system and allows one to remove entropy
in order to cool the system. Note that while optical pumping may lead to heating of the motional degrees of freedom, it is possible to recool the control atom afterwards, e.g., by sideband cooling. The two qubit gate $U_{\rs Z,i}(\theta)$
is implemented in close analogy to the many-body Rydberg gate $U_{g}$
previously discussed. For small phases $\theta$ the operator $\Sigma$ can be expanded, and the density matrix $\rho$ of the spin system evolves in one dissipative time step according to the Lindblad form
\begin{equation}
\partial_{t}\rho=\kappa\left[c_{p}\rho c_{p}^{\dag}-\frac{1}{2}\left\{ c_{p}^{\dag}c_{p}\rho+\rho c_{p}^{\dag}c_{p}\right\} \right]+o(\theta^{3})
\end{equation}
with the jump operators $c_{p}$ given in Eq.~(\ref{eq:Kitaev_jump_operators})
and the cooling rate $\kappa=\theta^{2}/\tau$. 
Note, that the cooling also works for large phases $\theta$, and therefore the most efficient dissipative state preparation 
is achieved with $\theta=\pi$.

The above scheme for the implementation of the many-body interaction
$A_{p}$ and the dissipative cooling with $c_{p}$ can be naturally
extended to arbitrary many-body interactions between the system spins
surrounding the control atom, as e.g., the $B_{p}$ interaction terms in the above toric
code. Gate operations on single system spins allow
to transform $\sigma_{i}^{x}$ in $\sigma_{i}^{y}$ and $\sigma_{i}^{z}$, in accordance with previous proposals for digital simulation of spin Hamiltonians \cite{Sorenson1999},
while selecting only certain spins to participate in the many-body gate
via local addressability gives rise to the identity operator for the non-participating spins. Consequently, we immediately obtain the implementation
of the general many-body interaction and jump operators 
\begin{equation}
A_{\alpha}=\prod_{i}W_{i},\hspace{30pt}c_{\beta}=\frac{1}{2}Q_{i}\left[1-\prod_{j}W_{j}\right]
\label{toolbox}
\end{equation}
with $W_{i},Q_{i}\in\{1,\sigma_{i}^{x},\sigma_{i}^{y},\sigma_{i}^{z}\}$.
Here, $\alpha$ and $\beta$ stand for a collection of indices characterizing
the position of the local interaction and the interaction type. 
Note
that $A_{\alpha}$ also includes single particle terms, as well as
two-body interactions.

\section{Toolbox for digital quantum simulation}

Extending the analysis to  a \textit{large lattice system}
with different, possibly non-commuting interaction terms in the Hamiltonian, i.e., $H=\sum_{\alpha}E_{\alpha}A_{\alpha}$
and dissipative dynamics described by a set of jump operators $c_{\beta}$
with damping rates $\kappa_{\beta}$, provides a complete toolbox for the quantum simulation of
many-body systems. Each term is characterized by a phase $\phi_{\alpha}$ ($\theta_{\beta}$)
written during a single time step determining its coupling energy $E_{\alpha}=\hbar\phi_{\alpha}/\tau$ and 
damping rate $\kappa_{\beta}=\theta_{\beta}^{2}/\tau$.  For small phases $\phi_{\alpha}\ll1$
and $\theta_{\beta}^{2}\ll1$, the sequential application of the gate
operations for all interaction and damping terms reduces to the master
equation of Lindblad form,
\begin{equation}
\partial_{t}\rho=-\frac{i}{\hbar}\left[H,\rho\right]+\sum_{\beta}\kappa_{\beta}\left[c_{\beta}\rho c_{\beta}^{\dag}-\frac{1}{2}\left(c_{\beta}^{\dag}c_{\beta}\rho+\rho c_{\beta}^{\dag}c_{\beta}\right) \right].
\end{equation}
The choice of the different phases during each time step allows for the control of the relative interaction strength of the different terms, as well as the simulation of inhomogeneous and time dependent systems.

The characteristic energy scale for the interactions $E_{\alpha}$ and damping rates $\kappa_{\beta}$ are determined
by the ratio between the time scale $\tau$ required to perform a single time step, and the phase difference $\phi_{\alpha}$ and $\theta_{\beta}$ 
written during these time steps. It is important to stress that within our setup, the interactions are quasi-local  and only 
influence the spins surrounding the control qubit. Consequently, the lattice system can be divided into a set of sublattices on which all gate operations that are needed for a single time step $\tau$, can be carried out in parallel. Then, the time scale for a single step $\tau$ becomes 
independent on the system size and is determined by the product of the number $z$ of such sublattices and the duration $\tau_s$ of all 
gate operations on one sublattice. In our setup, $\tau_s$ is mainly limited by the duration of the many-body Rydberg gate $U_g$, which 
is on the order of  $ \sim 1 \mu{\rm s}$ (see Methods section for details). For the toric code discussed above, we have to apply the many-body gate 
twice for every interaction term (see Fig. \ref{fig2}), and with $z=4$, we obtain $\tau \sim$ a few $\mu$s, resulting in characteristic energy scales and cooling rates 
of the order of hundred kHz. For the simulation of Hamiltonian dynamics this energy scale may be somewhat lower if Trotterization errors have to be taken care of.  It is a crucial aspect of this quantum simulation with Rydberg atoms that it can be performed fast and is compatible with current experimental time scales of cold atomic gases \cite{Bloch2008}.

Finally, we would like to point out that imperfect gate operations provide in leading order small perturbations for the Hamiltonian dynamics and
weak dissipative terms; see Methods section  and Fig.~\ref{fig3} for a numerical analysis of the induced errors. 
However, the thermodynamic properties and dynamical behaviour of a strongly interacting many-body system are in general robust to small 
perturbations in the Hamiltonian; e.g., the stability of the toric code for small magnetic fields
has  recently been demonstrated \cite{Vidal2009}.
 Consequently, small imperfections in the implementation of the gate operations
are tolerable.

An important aspect for the characterization of the final
state is the measurement of correlation functions $\chi=\langle A_{\alpha_{1}}\ldots A_{\alpha_{n}}\rangle$, where $A_{\alpha_{j}}$ 
denote local, mutually commuting many-body observables. In our scheme, the observables $A_{\alpha_{j}}$ can be measured via 
the mapping $G$ of the system information onto auxiliary qubits and their subsequent state selective detection. In analogy to noise 
correlation measurements in cold atomic gases \cite{Folling2005,Altman2004} the repeated measurement via such a detection 
scheme provides the full distribution function for the observables, and therefore allows to determine the correlation function $\chi$ 
in the system. Consequently, in the above discussion of Kitaev's toric code, the necessary string operators characterizing topological 
order can be detected.

\section{Lattice Gauge Theory}

In the first example, we discussed the implementation of the quantum simulator for the toric code,
and the extension to more complex stabilizer states is
straightforward. In the following, we will show that our approach can also be extended to systems 
with non-commuting terms in the Hamiltonian.
As an example, 
we focus on a three-dimensional $U(1)$-lattice gauge theory \cite{Kogut1979}, and show that dissipative ground state cooling can also
be achieved in such complex models. 
Such models have attracted a lot of recent interest in the search for `exotic' phases  and spin liquids 
\cite{Moessner2001,Motrunich2002,Hermele2004,Levin2005}. 
 The three-dimensional setup consists of spins located on the links of a cubic lattice (see Fig. \ref{fig1}d). The lattice structure for the spins
can be viewed as a corner sharing lattice of octahedra with one site of the cubic lattice in the center of each octahedra.
The Hamiltonian for the $U(1)$ lattice gauge theory takes the form
\begin{equation}
 H =U \sum_{o}\left( S^{z}_{o}\right)^{2}   -  J\sum_{p}B_{p} + V \: N_{\rs RK},
 \label{ringexchange}
\end{equation}
where the first term  in the Hamiltonian defines a low energy sector consisting of allowed spin configuration  with
an equal number of up and down spins on each octahedron, i.e., spin configurations with vanishing total spin 
$S_{o}^{z} = \sum_{i \in o} \sigma^{z}_{i}$ on each octahedron. The second term denotes a ring exchange interaction on 
each plaquette with $B_{p}=S_{1}^{+}S_{2}^{-}S_{3}^{+}S_{4}^{-}+S_{1}^{-}S_{2}^{+}S_{3}^{-}S_{4}^{+}$; 
here $S_{i}^{\pm}=\left[\sigma_{i}^{x}\pm i\sigma_{i}^{y}\right]/2$ and the numbering is clockwise around the plaquette. 
This term flips a state with alternating up and down spins
on a plaquette, i.e., $|\uparrow,\downarrow,\uparrow, \downarrow\rangle_{p} \rightarrow |\downarrow,\uparrow, \downarrow,\uparrow\rangle_{p}$.
The last term denotes the the so-called Rokhsar-Kivelson term, which counts the total number of flipable plaquettes $N_{\rs RK} = \sum_{p} B_{p}^2$.
While the ring exchange interaction commutes with the spin constraint, ring exchange terms on neighboring 
plaquettes are non-commuting. At the Rokhsar-Kivelson point with $J=V$, the system becomes exactly solvable \cite{Rokhsar1988}, 
and it has been proposed that in the regime 
$0\leq V\leq J$ the ground state is determined by a spin liquid smoothly connected to the Rokhsar-Kivelson point \cite{Hermele2004}:
the properties of this spin liquid are given by an artificial `photon' mode, gapped excitations
carrying  an `electric' charge (violation of the constraint on an octahedron), which interact with a 1/r 
Coulomb potential mediated by the artificial photons, and gapped magnetic monopoles.

\begin{figure*}[h!]
\includegraphics[width=1\columnwidth]{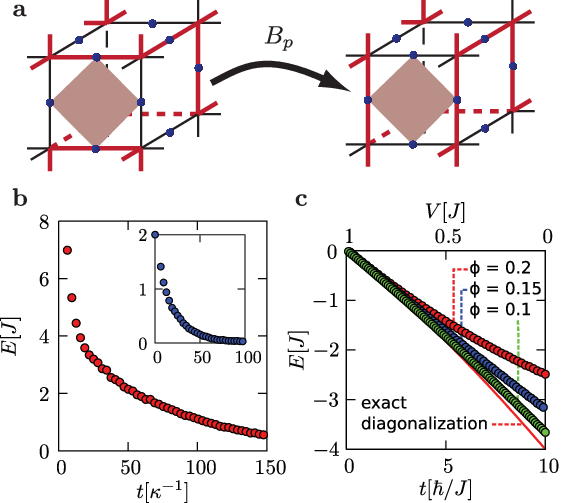} 

\caption{Lattice gauge theory: a) Illustration of a dimer covering with three dimers (red links) meeting at each site of the
cubic lattice. The front plaquette represents a flipable plaquette, which is transformed under the action of the ring-exchange $B_{p}$ 
into a different dimer covering.
b) Numerical simulation for the cooling into the ground state at the Rokhsar-Kivelson point with $E=0$ for a system with 4 unit cells (12 spins). 
The cooling into the constraint on the octahedra follows in analogy to the cooling of the toric code via the diffusion and annihilation of 'electric charges' on the octahedra.
The inset shows the cooling into the equal superposition of all dimer coverings starting from an initial state satisfying the constraint on all octahedra.
c) Coherent time evolution from the Rokhsar-Kivelson point with a linear ramp of the Rokhsar-Kivelson term $V(t) = J( 1-t J /10\hbar)$: the solid line denotes
the exact ground state energy, while the dots represent the digital time evolution during an adiabatic ramp for different phases $\phi$ written during each time step.
The difference accounts for errors induced by the Trotter expansion due to the non-commutative terms in the Hamiltonian.  }
\label{fig4} 
\end{figure*}

In the following, we present the implementation of this Hamiltonian within our scheme for the digital 
quantum simulation and demonstrate that dissipative ground state cooling can be achieved at the 
Rokhsar-Kivelson point. The control qubits reside in the center of each octahedron (on the lattice sites 
of the 3D cubic lattice) controlling the interaction on each octahedron, and in the center of each plaquette 
for the ring exchange interaction $B_{p}$, see Fig.~\ref{fig1}. Then, the coherent 
time evolution of the Hamiltonian (\ref{ringexchange}) can be implemented in analogy to the above discussion 
by noting that the ring exchange interaction $B_{p}$ and $N_{\rs RK}$ can be written as a sum of four-body 
interactions of the form (\ref{toolbox}), while the constraint on the octahedra is an Ising interaction, 
see Methods section. Next, we discuss the jump operators for the dissipative ground state preparation.
The cooling into the subspace with an equal number of up and down spins on each octahedron
is obtained by the jump operator
\begin{equation}
c_{s}=\frac{1}{4}\left[1+\prod_{j}e^{-i\frac{\pi}{6}\sigma_{j}^{z}}\right]\sigma_{i}^{x}\left[1-\prod_{j}e^{i\frac{\pi}{6}\sigma_{j}^{z}}\right],
\label{eq:lattice_gauge_jump_operators}
\end{equation}
where the product is carried out over the six spins located on the
corners of the octahedron (see Fig. \ref{fig1}d). The  {}``interrogation'' part $1-\prod_{j}\exp({i\frac{\pi}{6}\sigma_{j}^{z}})$ of the jump
operator vanishes if applied to any state with three up and three down spins, while in all other cases a spin
is flipped. Then the cooling follows in analogy to the cooling in the toric code by the diffusion of the `electric' charges.  
Identifying each spin up with a `dimer' on the link, all states satisfying the constraints on the octahedra can be viewed as
a dimer covering with three dimers meeting at each site of the cubic lattice, see Fig.~\ref{fig4}a. Within this description, 
the ground state at the Rokhsar-Kivelson point is given by the condensation of the dimer coverings \cite{Levin2005}, i.e., 
the equal weight superposition of all dimer coverings.
The condensation of the dimer coverings is then achieved by the jump operator
\begin{equation}
  c_{p} = \frac{1}{2}\sigma_{i}^{z}\left[1-B_{p}\right]B_{p}.
\end{equation}
This jump operator has two dark states, which are the $0$ and $+1$ eigenstates of $B_p$. The $0$ eigenstate 
corresponds to a non-flippable plaquette, while the $+1$ eigenstate is the equal weight superposition of the 
original dimer covering and the dimer covering obtained by flipping the plaquette (i.e., the $+1$ eigenstate). 
Finally, the jump operator $c_p$ transforms the third eigenstate with eigenvalue $-1$ into the $+1$ eigenstate.
After acting on all plaquettes, the system is cooled into the dark
state which is the equal superposition of all dimer coverings, which can be reached by flipping different plaquettes. 
The cooling of these jump operators is demonstrated via a numerical simulation for a small system of 4 unit cells, see Fig.~\ref{fig4}b.

The implementation of the digital quantum simulations provides full control on the spatial and temporal
interaction strengths. Therefore, there are two possibilities to analyze the phase diagram for arbitrary interaction 
strengths: (i) The possibility to vary the different coupling strengths in time allows us adiabatically explore the phase diagram;
the adiabatic preparation using the Trotter expansion is shown in Fig.~\ref{fig4}c.
(ii) On the other hand,  the spatial control of the coupling parameters allows us to divide the lattice into a system 
and a bath. The ground state of the bath is given by the Rokhsar-Kivelson state, which can be continuously cooled
via the dissipative terms,  while the system part is sympathetically cooled due to
its contact with the bath; in analogy to the cooling well known in
condensed matter systems.

\section{Methods}

\subsection{Gate errors}

In the following, we discuss the influence of a gate error onto the
dynamics of the system. For simplicity, we illustrate the general
behaviour for an error in the many-body gate $U_{g}$ for the coherent
time evolution of the many-body interaction $A_{p}$. 
The imperfect many-body gate operation can be written \begin{equation}
\tilde{U}_{g}=|0\rangle\langle0|_{c}\otimes e^{i\phi Q}+|1\rangle\langle1|_{c}\otimes A_{p},\label{gateerror}\end{equation}
 where the perfect gate $U_{g}$ is recovered for $Q\rightarrow0$
and the operator $Q=Q^{\dag}$ acts on the system spins surrounding
the control atom. This form of the error is motivated by the specific
implementation of the gate \cite{Muller2009}; however, it can be
seen that different errors in the many-body and single particle gates
will lead to similar phenomena.  For the \textit{coherent} time evolution, the imperfect
gate gives rise to a finite amplitude for the control qubit to end
up in the state $|1\rangle_{c}$. Consequently, optical pumping of
the control qubit is required to reinitialize the system. Then the
gate operations on a single plaquette give rise to the mapping of
the density matrix onto \begin{equation}
\rho\rightarrow C\rho C^{\dag}+D\rho D^{\dag}\end{equation}
 with ($\Theta\equiv e^{i\phi Q}$) \begin{eqnarray}
C & = & \frac{1}{2}\left[\left(\Theta^{2}+{\bf 1}\right)\cos\phi+\left(\Theta A_{p}+A_{p}\Theta\right)i\sin\phi\right]\\
 & \approx & \exp\left[i\phi\left(A_{p}+Q\right)\right]-\frac{\phi^{2}}{2}Q^{2}\\
D & = & \frac{1}{2}\left[\left(\Theta^{2}-{\bf 1}\right)\cos\phi+\left(\Theta A_{p}-A_{p}\Theta\right)i\sin\phi\right]\\
 & \approx & -i\phi Q.\end{eqnarray}
 The last equations hold with an accuracy up to third order in the
small parameter $\phi$. Consequently, the optical pumping has no
influence in leading order, and the system is well described by a
Hamiltonian evolution with the modified Hamiltonian $H=-(\hbar/\tau)\phi\left[A_{P}+Q\right]$.
The characteristic energy scale of the correction is again given by
$\hbar\phi/\tau$, and consequently describes a small perturbation
if $|Q|\ll1$. In second order expansion in the small parameter $\phi$,
the mapping of the density matrix reduces to \begin{equation}
\rho\rightarrow\rho-i\phi\left[h,\rho\right]-\frac{\phi^{2}}{2}\left[h,\left[h,\rho\right]\right]+\frac{\phi^{2}}{2}\left(2Q\rho Q-\left\{ Q^{2},\rho\right\} \right),\end{equation}
 with $h=-[A_{p}+Q]$. The first terms on the right hand side describe
the coherent evolution of the system with the evolution operator $\exp(-i\phi h)$
consistently expanded up to second order, while the last term takes
the standard Lindblad form for a dissipative coupling with the jump
operator $c_{e}=Q$ describing a dephasing with the rate $\kappa_{e}=\phi^{2}/\tau$.

\subsection{Mesoscopic Rydberg gate}

\label{app:gate}

In the following we briefly summarise the main properties and requirements
of the many-body Rydberg gate $U_{g}$ introduced in Ref.~\cite{Muller2009}.
The internal level structure of the control atom and the surrounding
ensemble atoms is depicted in Fig.~\ref{fig1}. The underlying physical
mechanism of the gate operation (\ref{eq:gate}) is a conditional
Raman transfer of all ensemble atoms between their logical internal
states $|A\rangle$ and $|B\rangle$, which - depending on the internal
state $|0\rangle$ or $|1\rangle$ of the control qubit - is either
inhibited or enabled. The gate is realised by the following three
laser pulses: (i) A first state selective $\pi$-pulse acting on the
control atom changes the ground state $|1\rangle$ into the Rydberg
state $|r\rangle$. (ii) During the whole gate operation, a strong
coupling laser of Rabi frequency $\Omega_{c}$ constantly acts on
all ensemble atoms and off-resonantly couples the Rydberg level $|R\rangle$
to the intermediate level $|P\rangle$ with a detuning $\Delta$.
Its frequency is chosen such that it is in two-photon resonance with
the two Raman laser beams of Rabi frequency $\Omega_{p}$ (see Fig.~\ref{fig1}),
thereby establishing a condition known as electromagnetically induced
transparency (EIT) \cite{Fleischhauer2005}. The system then adiabatically follows a zero energy dark state, which at the end of the pulse is identical to the one in the beginning. In consequence, when 
the Raman laser pulses are applied - and provided the control atom
resides in state $|0\rangle$ - this two-photon resonance condition effectively blocks the Raman transfer from $\ket{A}$ to $\ket{B}$. In case the control atom was excited
to the Rydberg state $|r\rangle$ in step (i), the large Rydberg-Rydberg
interaction energy shift (dipole blockade) lifts the blocking condition
for the ensemble atoms and thus the Raman transfer takes place. (iii)
Finally, the control atom is transfered from state $|r\rangle$ back
to $|1\rangle$ via a second $\pi$-pulse. The total time $T_{\textup{gate}}$
required for the gate is mainly limited by the duration of the Raman
pulse, resulting in $T_{\textup{gate}}\sim16\pi\Delta/(3\Omega_{p}^{2})$.
The principal error source for the many-body gate is due to the Rydberg-Rydberg interactions between the ensemble atoms. For $N$ ensemble atoms the accumulated phase errors scale like $\phi|Q| \sim N(N-1)(\Omega_p/\Omega_c)^2$, i.e., they depend on all possible pair combinations and the probability of an ensemble atom being excited to the Rydberg state due to non-adiabatic processes \cite{Muller2009}.

\subsection{Experimental implementation}

Our setup consists of control and ensemble atoms  trapped in large
spacing optical lattices (see Fig.~1), so that single site addressability can be achieved. In order to manipulate ensemble atoms independently, their spacing $a$ must be larger than the wavelength $\lambda_p$ of the Raman lasers for the many-body gate. Such a spatial resolution can be achieved by tightly focussing the laser beam, by employing superlattice beams for the gate pulses, or sub-wavelength addressing techniques based on magnetic field gradients \cite{Stokes1991} or dark state resonances \cite{Gorshkov2008}. Control and ensemble atoms can be distinguished spectroscopically, e.g., by using different hyperfine states in two state-dependent lattices.  A suitable set of parameters is determined by balancing the need for sufficiently large lattice spacing with at the same time strong Rydberg interactions for a fast and high-fidelity many-body gate.

We require the ensemble atoms to be separated by $a = 3.5\lambda_p \approx 1.5\,\mu{\mathrm{m}}$ and a fast many-body gate with $T_{\mathrm{gate}}=1.5\,\mu\mathrm{s}$ (which is much shorter than decoherence times, e.g., due to radiative decay of the Rydberg states). For $^{87}\mathrm{Rb}$ this is achieved by choosing $\Omega_{p}=2\pi\times67\,\mathrm{MHz}$,
$\Delta=2\pi\times2\,\mathrm{GHz}$, $\Omega_{c}=2\pi\times1\,\mathrm{GHz}$, an interaction strength of $V = 10\hbar\Omega_c^2/\Delta$, and using the Rydberg states $\ket{r}_c = \ket{59s}$ and $\ket{R}_i = \ket{53s}$, respectively.
Note that for these Rydberg states the corresponding distances are still larger than the LeRoy radius, i.e., there is no overlap between the wavefunctions of the atoms. Furthermore, corrections to a pure van der Waals interaction due to resonant dipole-dipole couplings \cite{Li2005,Walker2008} are also small, while using $s$ states ensures the interaction being isotropic.
For these parameters, the errors due to ensemble-ensemble interactions
result for $N = 4$ atoms in $\phi|Q|=0.2$.

Further errors arise from crosstalk between plaquettes being processed in parallel, i.e.~when a control atom interacts with ensemble atoms of distant plaquettes.  Due to the rapid decay of the van der Waals interaction the residual interaction is reduced by a factor of at least $125$ on a square lattice. For $V = 10\hbar\Omega_c^2/\Delta$ the resulting error is of similar size as due to ensemble-ensemble interactions. This error can be further reduced by increasing the number of sublattices $z$ such that only every second or third plaquette is processed in parallel.

\subsection{Quantum gates for the $U(1)$ lattice gauge theory}

The Hamiltonian giving rise to the constraint for the spins on the
octahedra can be expressed as a sum of Ising interactions,
\begin{equation}
  \left(S_o^z\right)^2 = \sum\limits_{i\ne j}^6\sigma_i^z\sigma_j^z + \mathrm{const},
\end{equation}
which allow for an efficient implementation using the general toolbox
for quantum simulation. The implementation for the jump operators for
the constraint is obtained in analogy to the general jump operators
with the many-body gate $U_{g}$ replaced by the gate
$|0\rangle\langle0|_{c}\otimes{\bf
  1}+|1\rangle\langle1|_{c}\otimes\prod_{i}\exp(i\frac{\pi}{6}\sigma_{i}^{z})$.
On the other hand, the ring exchange interaction can be written as a
sum of commuting four-body interactions \begin{eqnarray}
  B_{p}&=&\frac{1}{8}\sum\limits _{j=1}^{8}B_{p}^{(j)} =
  \frac{1}{8}(\sigma_{1}^{x}\sigma_{2}^{x}\sigma_{3}^{x}\sigma_{4}^{x}+\sigma_{1}^{y}\sigma_{2}^{y}\sigma_{3}^{y}\sigma_{4}^{y}\nonumber\\&&+\sigma_{1}^{x}\sigma_{2}^{x}\sigma_{3}^{y}\sigma_{4}^{y}
  +\sigma_{1}^{y}\sigma_{2}^{y}\sigma_{3}^{x}\sigma_{4}^{x}-\sigma_{1}^{x}\sigma_{2}^{y}\sigma_{3}^{x}\sigma_{4}^{y}\nonumber\\&&-\sigma_{1}^{y}\sigma_{2}^{x}\sigma_{3}^{y}\sigma_{4}^{x}+\sigma_{1}^{x}\sigma_{2}^{y}\sigma_{3}^{y}\sigma_{4}^{x}+\sigma_{1}^{y}\sigma_{2}^{x}\sigma_{3}^{x}\sigma_{4}^{y}).\end{eqnarray}
Likewise, the Rokhsar-Kivelson term can be decomposed into
\begin{eqnarray}
  B_{p}^2 &=& \frac{1}{8}\sum\limits _{j=1}^{8} N_{p}^{(j)} = \frac{1}{8}(\sigma_{1}^{0}\sigma_{2}^{0}\sigma_{3}^{0}\sigma_{4}^{0}-\sigma_{1}^{0}\sigma_{2}^{0}\sigma_{3}^{z}\sigma_{4}^{z}\nonumber \\&&+\sigma_{1}^{0}\sigma_{2}^{z}\sigma_{3}^{0}\sigma_{4}^{z}
 -\sigma_{1}^{0}\sigma_{2}^{z}\sigma_{3}^{z}\sigma_{4}^{0}-\sigma_{1}^{z}\sigma_{2}^{0}\sigma_{3}^{0}\sigma_{4}^{z}\nonumber \\&&+\sigma_{1}^{z}\sigma_{2}^{0}\sigma_{3}^{z}\sigma_{4}^{0}
  -\sigma_{1}^{z}\sigma_{2}^{z}\sigma_{3}^{0}\sigma_{4}^{0}+\sigma_{1}^{z}\sigma_{2}^{z}\sigma_{3}^{z}\sigma_{4}^{z}),
\end{eqnarray}
where $\sigma_i^0$ is the identity matrix. Consequently, the coherent
time evolution follows again from the general toolbox, while the jump
operators for cooling into the ground state at the Rokhsar-Kivelson
point effectively cool into the zero eigenvalue eigenstate of the operators
\begin{equation}
\frac{1}{2}[1-B_{p}]B_{p} =\frac{1}{16}\sum\limits_{j=1}^{16}C_p^{(j)} = \frac{1}{16}\left[\sum\limits_{j=1}^8B_p^{(j)}-\sum\limits_{j=1}^8N_p^{(j)}\right].
\end{equation}
This can be achieved by replacing the gate $U_g$ with
\begin{eqnarray}
  U_{B} & = & |0\rangle\langle0|_{c}\otimes{\bf
    1}+|1\rangle\langle1|_{c}\otimes\exp\left[i\frac{\pi}{2}(1-B_p)B_p\right]\nonumber\\ &=&
  \prod_{j=1}^{16} U_{c}(\pi/2)^{-1}U_jU_{c}(\pi/2)\exp\left(i\pi/32\sigma_{c}^{z}\right)\nonumber\\&&\phantom{\prod}U_{c}(\pi/2)^{-1} U_jU_{c}(\pi/2),
\end{eqnarray}
with $U_j = |0\rangle\langle0|_{c}\otimes{\bf
  1}+|1\rangle\langle1|_{c}\otimes C_p^{(j)}$. This gate operation
leaves states with eigenvalue $0, +1$ of $B_p$ invariant, while the
$-1$ eigenvalue picks up a phase of $\pi$. It can be implemented as a
product of many-body gates which derive directly from the standard
gate $U_{g}$ with the combination of spin rotations.

\subsection{Acknowledgments}

This work was supported by the Austrian Science Foundation (FWF),
and by the Deutsche Forschungsgemeinschaft (DFG) within SFB/TRR 21.

\subsection{Author contributions}

All five authors contributed equally to all parts of this work.

\subsection{Author information}
Reprints and permissions information is available online at http://npg.nature.com/reprintsandpermissions.
Correspondence and requests for materials should be addressed to H.W.

\clearpage

\setcounter{page}{1}


\end{document}